\documentstyle{article}

\title{\bf The 2nd order corrections to the interaction of two
reggeized gluons from the bootstrap}
\author{Mikhail Braun$^{a)}$, and Gian Paolo Vacca$^{b,c)}$  \\
$^{a)}$ Department of high-energy physics, University of S. Petersburg\\
$^{b)}$Department of Physics, University of Bologna\\
$^{c)}$II. Institute f\"ur Theoretische Physik, Universit\"at Hamburg.}
\def\beq{\begin{equation}}
\def\eeq{\end{equation}}
\def\noi{\noindent}
\def\oq{\omega(q)}
\def\oa{\omega(q_{1})}
\def\ob{\omega(q_{2})}
\def\eq{\eta (q)}
\def\ea{\eta (q_{1})}
\def\eb{\eta (q_{2})}
\def\ec{\eta (q'_{1})}
\def\ed{\eta (q'_{2})}
\def\ee{\eta(q_1-q'_1)}
\def\xq{\xi(q)}
\def\xa{\xi(q_1)}
\def\xb{\xi(q_2)}
\def\xc{\xi(q'_1)}
\def\xd{\xi(q'_2)}
\def\xe{\xi(q_{11'})}

\def\Ga{{\rm\Gamma}}
\def\ps{{\rm\psi}}
\begin{document}
\maketitle
\medskip
\noi{\bf Abstract.}
The 2nd order corrections are obtained 
to both forward and nonforward 
interaction of reggeized gluons in the octet colour channel
using as a basis the
bootstrap relation and a specific ansatz to solve it.
 The obtained forward kernel
coincides with the logarithmic term plus two first
non-logarithmic terms in the Pomeron 2nd order kernel.
Both forward and nonforward kernels are found to
be infrared finite.

\newpage
\section{ Introduction.}
Recently the 2nd order corrections were calculated for the forward
hard (BFKL) pomeron equation [1,2]. They clearly divide into a
correction for the running of the coupling and the 2nd order
corrections in the running coupling proper. Generalization of
this result for the other physically important case, that of the
odderon, requires knowledge of the 2nd order corrections to the
non-forward interaction of two reggeized gluons in the octet colour
state. It is well-known that this interaction is severely restricted
by the so-called bootstrap relation, which is, in fact, the
unitarity requirement for the octet channel [3]. The form this
relation takes in the 2nd order was recently discussed in [4].

In  previous publications one of the authors (MAB) proposed to
use the bootstrap relation and a specific ansatz to satisfy it
for the introduction of the running
constant, to all orders of the fixed one, into the gluon octet
interaction [5]. 
Since in the lowest order this interaction is just one half of
the interaction in the vacuum (pomeron) channel, this method also
serves to find corrections for the running coupling in the 
pomeron equation both for forward and non-forward cases.
Note that whereas for the forward case these corrections are trivial 
and can easily be recostructed via the renormalization group, this
is not so for the non-forward case, with more than one scale.

In this paper, using the bootstrap and the ansatz of [5], we calculate the 
full second order
 corrections for the non-forward gluon interaction in the octet channel
necessary for the odderon equation. In this way we  also automatically find
the corrections for the
running of the coupling for the non-forward pomeron equation.

\section{Basic equations}
Up to the 2nd order in the fixed coupling, two gluons are described by 
a Schr\"odinger equation in the transverse space
\beq
H\phi(q_1,q_2)=E\phi(q_1,q_2).
\eeq
Here $q_1$ and $q_2$ are the momenta of the two gluons; $E=1-j$ where
$j$ is the angular momentum. The Hamiltonian $H$ is a sum of the
kinetic energy, given by a sum of gluonic Regge trajectories with 
a minus sign
$-\oa-\ob$, and the interaction $-V^{(R)}$, which depends on the colour
state $R$ of the two gluons. We shall be interested in either the
vacuum state ($R=1$) or the octet state ($R=8$). As mentioned, to the
lowest order in the running coupling, $V^{(1)}=2V^{(8)}$.

The bootstrap relation requires that the solution to Eq. (1) for the 
octet channel be the reggeized gluon itself. This can be fulfilled if
the interaction $V_8$ is related to the trajectory as
\beq
\int (d^{2}q'_{1}/(2\pi)^{2})V^{(8)}(q,q_{1},q'_{1})=
\oq-\oa-\ob,\ \ q=q_1+q_2.
\eeq
In [5] it was shown that this relation can be satisfied provided
both $V^{(8)}$ and $\omega$ are expressed via a single function $\eta(q)$
as follows
\beq
\oq=-\int
\frac{d^{2}q_{1}\eq}{(2\pi)^{2}\ea\eb},\ \ 
q=q_{1}+q_{2}
\eeq
and
\beq
V^{(8)}(q,q_1,q'_{1})=
\left((\frac{\ea}{\ec}+\frac{\eb}{\ed})
\frac{1}{\eta(q_1-q'_1)}-\frac{\eq}{\ec\ed}\right).
\eeq
In the lowest order of the fixed coupling one has (we take $N_c=3$)
\beq
\eta^{(0)}(q)=\frac{q^2}{3\alpha_s}.
\eeq

The forms (3) and (4) guarantee that to the lowest order in the running
coupling the full interaction in the vacuum channel is infrared
stable [5]. A stronger result is that they also guarantee that the sum of 
single and
pair terms in the odderon equation stays infrared stable in all
orders in the running coupling. Indeed, this sum can be represented as
a sum of three pair terms $-(1/2)(K_{12}+K_{ 23}+K_{ 31})$, where e.g.
\beq
K_{12}=\oa+\ob +2V^{(8)}_{12},
\eeq
and this combination is infrared stable provided $\eq$ goes to zero
as $q^2\rightarrow 0$ modulo logarithms. 
Of course, this does not settle the question of the eventual infrared
stability of the odderon equation in the higher orders, since already
at the 2nd order triple interaction terms seem to appear. We shall not
deal with this problem here.

Our aim will be to study the 2nd order corrections to the kernel $K$
which follow from the 2nd order corrections in $\eq$. The latter can be
found from the known form of the trajectory $\oq$ to the 2nd order in
the fixed coupling [6]:
\beq
\oq=\omega^{(1)}(q)+\omega^{(2)}(q),
\eeq
where the superscripts refer to the order in the coupling, and,
in the dimensional regularization in the $\overline{{\rm MS}}$ scheme,
\beq
\omega^{(1)}(q)=-\bar{g}^2\left(\frac{2}{\epsilon}+2t+
\epsilon(t^2-2\psi'(1))\right),
\eeq
\beq
\omega^{(2)}(q)=-\bar{g}^4\left(A(\frac{1}{\epsilon^2}-t^2)+
B(\frac{1}{\epsilon}+2t)+C\right),
\eeq
Here and in the following $t=\ln (q^2/\mu^2)$; 
$\mu$ is the normalization point; 
$\epsilon\rightarrow 0$;
\beq
\bar{g}^2=\frac{3\alpha_s{\rm \Gamma}(1-\epsilon)}{(4\pi)^{1+\epsilon}}
\eeq
and $\bar{g}^4$ is just the square of (10). The coefficients $A,B$ and $C$ are
\beq
A=\frac{11}{3}-\frac{2}{9}N_f;\ B=\frac{67}{9}-\frac{\pi^2}{3}-
\frac{10}{27}N_f;\ C=-\frac{404}{27}+2\zeta(3)+\frac{56}{81}N_f.
\eeq
We have retained a term proportional to $\epsilon$ in (8) for future
reference.
\section{Function $\eq$ to the 2nd order}
We present function $\eq$ in the form
\beq
\eq=\eta^{(0)}(q)(1+\xi(q)+...),
\eeq
where $\xi$ represents the terms of the 2nd order in $\alpha_s$.
The substitution of the running coupling
instead of the fixed one into (5) evidently gives a contribution to $\xi$
\[-\frac{3\alpha_sA}{4\pi}t.\]
This term is what we call a correction for the running of the coupling.
However there may be other terms in $\xi$ which represent corrections
of the 2nd order in the already running coupling constant.
We are going to find all these by matching the form of $\omega^{(2)}$ found
via function $\eq$ given by (12) with the expression (9) found from the
perturbative QCD.

Putting (12) into Eq. (3) we obtain in the 2nd order
\beq
\omega^{(2)}(q)=\omega^{(1)}(q)\xi(q)+\frac{6\alpha_s}{4\pi^2}
\int\frac{d^2q_1 q^2}{q_1^2q_2^2}\xi(q_1).
\eeq
With $\omega^{(2)}(q)$ known, this is an integral equation for $\xi$.
We shall solve it  by choosing an appropriate functional form of $\xi(q)$:
\beq
\xi(q)=c+dt+\epsilon ft^2,
\eeq
where the coefficients $c,d$ and $f$ in principle may depend on $\epsilon$
but are regular at $\epsilon=0$. We introduce an extra factor $\epsilon$
into the third term of (14) to avoid poles of the third order in $\epsilon$
absent both in $\omega^{(2)}$ and the first term on the r.h.s. of (13).

We denote the integral term in (13) as $X$. In the dimensional regularization
we have
\beq
X=\frac{6\alpha_sq^2}{(2\pi)^{D}}(cX_0+dX_1+f\epsilon X_2),
\eeq
where
\beq
X_n=\int\frac{d^Dq_1t_1^n}{q_1^2q_2^2},
\eeq
$D=2+2\epsilon$ and $t_1=\ln(q_1^2/\mu^2)$. $X_n$ can be calculated as the $n$-th derivative
in $\alpha$, taken at $\alpha=0$ of a general integral
\beq
I=\int\frac{d^Dq_1q^{2\alpha}}{q_1^2q_2^2}.
\eeq
This integral can be easily calculated by standard methods using the
generalized Feynman representation. We find:
\beq
I=\pi^{1+\epsilon}(q^2)^{\alpha+\epsilon-1}
\frac{{\rm \Gamma}(1-\alpha-\epsilon)\Ga(\alpha+\epsilon)
\Ga(\epsilon)}{\Ga(1-\alpha)\Ga(\alpha+2\epsilon)}.
\eeq
>From this we obtain
\beq
X_0=\pi^{1+\epsilon}(q^2)^{\epsilon-1}
\frac{{\rm \Gamma}(1- \epsilon){\rm\Gamma}^2(\epsilon)}
{\Ga(2\epsilon)},
\eeq
\beq
X_1=X_0Y,\ \ Y=(t-\ps(1-\epsilon)+\ps(\epsilon)+\ps(1)-\ps(2\epsilon)),
\eeq
\beq
X_2=X_0(Y^2+\ps'(1-\epsilon)+\ps'(\epsilon)-\ps'(1)-\ps'(2\epsilon)).
\eeq
One has to take into account that the 1st order trajectory (8) is
related to $X_0$ as
\beq
\omega^{(1)}(q)=-\frac{3\alpha_s q^2}{(2\pi)^{2+2\epsilon}}X_0.
\eeq

Combining the two terms on the r.h.s of (13) we  get
\beq
\omega^{(2)}(q)=\frac{3\alpha_sq^2}{(2\pi)^{2+2\epsilon}}X_0
\left(-\frac{u}{\epsilon}+c+ht+f\epsilon t^2\right)
\eeq
with $u=d-2f$.
Recalling the form of $\omega^{(1)}$, Eq. (8), separating $\bar{g}^2$ and
neglecting terms of the order $\epsilon$ or higher, we finally get
\beq
\omega^{(2)}(q)=2\bar{g}^2\left(-\frac{u}{\epsilon^2}+
\frac{c}{\epsilon}+ct+\frac{1}{2}dt^2+u\ps'(1)\right).
\eeq

Comparing this expression with the one found from the perturbative QCD (9),
 we note
at once that to match them we have to assume that both $u$ and $c$ contain
terms of the higher order in $\epsilon$:
\beq
u=u_0+u_1\epsilon+u_2\epsilon^2,\ \ c=c_0+c_1\epsilon.
\eeq
With this form the 2nd order trajectory becomes
\beq
\omega^{(2)}(q)=2\bar{g}^2\left(-\frac{u_0}{\epsilon^2}+
\frac{c_0-u_1}{\epsilon}+c_0t+\frac{1}{2}dt^2+c_1-u_2+u_0\ps'(1)\right)
\eeq.
So we have just 5 parameters $c_0,c_1, d, u_0, u_1, u_2$ to match this
expression with  5 terms in (9) containing 
$1/\epsilon^2, 1/\epsilon, t, t^2$ and a constant.
We get for our parameters
\beq
u_0=(1/2)\bar{g}^2A,\
c_0-u_1=-(1/2)\bar{g}^2B,\
c_0=-\bar{g}^2B,\
d=\bar{g}^2A,\
u_0\ps'(1)+c_1-u_2=-(1/2)\bar{g}^2C.
\eeq
We observe that to match the perturbative trajectory it is not
sufficient to take into account only the corrections to $\eq$ due to
the running of the coupling, which  corresponds to all 
coefficients  equal to zero, except for equal $d$ and $u_0$.
Quite a few proper 2nd order corrections have to be also included.
\section{Forward interaction in the octet channel}
Having determined the 2nd order corrections for $\eq$ we are now in a 
position to find these corrections for the interaction of gluons in the
octet channel. As mentioned, to the lowest order in the running coupling,
the vacuum channel interaction is just twice the octet one. So we also
find the corrections due to the running of the coupling for the vacuum 
channel. In this section we calculate the 2nd order interaction for the 
forward case $q=0$, which is a much simpler task due to the fact that
the eigenfunctions of the 1st order kernel are simple.

As for the pomeron, at $q=0$ the eigenvalue equation for the octet channel
interaction
\beq
-K\phi=E\phi, 
\eeq
with $K$ given by (6),
can be simplified by substituting
\[ 
\phi(q)\rightarrow\eq\phi(q).\]
Then (28) takes the form
\beq
2\oq\psi(q)+\frac{4}{4\pi^2}\int\frac{d^2q'\phi(q')}{\eta(q-q')}=-E\psi(q).
\eeq
In the 1st order this is just the BFKL equation. In the 2nd order we get for
the left-hand side
\beq
K^{(2)}\psi\equiv 2\omega^{(2)}\psi(q)-\frac{12\alpha_s}{4\pi^2}
\int\frac{d^2q'\xi(q-q')\psi(q')}{(q-q')^2},
\eeq
where $\xi$ is defined by (12) and determined by (14) and (27).

We shall calculate the action of the kernel $K^{(2)}$ 
on the 1st order
BFKL eigenfunctions
\beq
\psi_\lambda(q)=q^{2\lambda}.
\eeq
In fact, for the normalizability one should take $\lambda=-1/2+i\nu$ and 
multiply (31) by the appropriate normalization factor, but this is
irrelevant for the properties of the kernel $K^{(2)}$.

We again introduce a notation $X$ for the integral part of the left-hand side
(without the minus sign). Putting (14) into the integrand we find the same 
form (15) for $X$, with an extra factor two and  $X_n$ which are now given by
\beq
X_n=\int\frac{d^Dq_1t_2^n}{q_1^{-2\lambda}q_2^2},
\ q_2=q-q_1,\ t_2= \ln (q_2^2/\mu^2).
\eeq
These integrals can again be obtained by differentiating in $\alpha$ at 
$\alpha=0$ the generic integral
\beq
I=\int\frac{d^Dq_1q_2^{2\alpha}}{q_1^{-2\lambda}q_2^2}
=\pi^{1+\epsilon}(q^2)^{\alpha+\epsilon-+\lambda}
\frac{{\rm \Gamma}(-\lambda-\alpha-\epsilon)\Ga(\alpha+\epsilon)
\Ga(1+\lambda+\epsilon)}{\Ga(1-\alpha)\Ga(-\lambda)\Ga(1+\alpha+
+\lambda+2\epsilon)}.
\eeq
>From this we find 
\beq
X_0=\pi^{1+\epsilon}(q^2)^{\epsilon+\lambda}
\frac{{\rm \Gamma}(-\lambda- \epsilon){\rm\Gamma}(\epsilon)
\Ga(1+\lambda+\epsilon)}
{\Ga(-\lambda)\Ga(1+\lambda+2\epsilon)},
\eeq
\beq
X_1=X_0Y,\ \ Y=t-\ps(-\lambda-\epsilon)+\ps(\epsilon)+\ps(1)-\ps(1+
\lambda+2\epsilon),
\eeq
\beq
X_2=X_0(Y^2+\ps'(-\lambda-\epsilon)+\ps'(\epsilon)-\ps'(1)-\ps'(1+
\lambda+2\epsilon)).
\eeq

Combining the terms we get for $X$
\[
X=\frac{12\alpha_s}{(2\pi)^{2+2\epsilon}}X_0
\Bigl[-\frac{u}{\epsilon}+c+u(t+\Delta)+\]\beq
\epsilon\Bigl(d(\ps'(-\lambda)+\ps'(1)-2\ps'(1+\lambda))+
f((t+\Delta)^2-\ps'(-\lambda)-2\ps'(1)+3\ps'(1+\lambda))\Bigr)\Bigr],
\eeq
where we denoted 
\beq
\Delta=2\ps(1)-\ps(-\lambda)-\ps(1+\lambda)
\eeq 
the eigenvalue of the 1st order kernel (up to a constant factor).
To obtain our final result we separate a factor to be included into
$\bar{g}^2$ and develop the rest, including $X_0$, in $\epsilon$ up to constant
terms. Then we get
\beq
X=4\bar{g}^2q^{2\lambda}\left( -\frac{u}{\epsilon^2}+\frac{c}{\epsilon}
+c(t+\Delta)+\frac{1}{2}d((t+\Delta)^2+\ps'(-\lambda)-\ps'(1+\lambda))+
u\ps'(1)\right).
\eeq

Now we have only to subtract it from the 1st term on the righthand 
side of (30).
Taking $\omega^{(2)}$ in the form (24) we find that the terms 
singular in $\epsilon$ cancel, so that the kernel turns out to be 
infrared stable, as
expected. Our final result for it is
\beq
K^{(2)}\psi=4\bar{g}^2\Bigl(-d\Delta t-c\Delta+\frac{1}{2}d(-\Delta^2+
\ps'(1+\lambda)-\ps'(-\lambda))\Bigr)\psi(q).
\eeq
One notes that out of five coefficients entering $\xi$ only two
remain in the final expression for the action of the kernel. Using (27) we
find
\beq
K^{(2)}\psi=
4(\frac{3\alpha_s}{4\pi})^2\Bigl(-\Delta At+B\Delta+\frac{1}{2}A
(-\Delta^2+\ps'(1+\lambda)-\ps'(-\lambda))\Bigr)\psi(q).
\eeq
The first term, proportional to $\ln (q^2/\mu^2)$ gives the correct
contribution from the running of the coupling. As mentioned this term
is the same for the octet and vacuum channels. So the bootstrap indeed
gives the correct description of the running of the coupling, at least, in
the 2nd order. Other terms are corrections to the interaction in the 2nd
order in the running coupling constant. They refer only to the octet
interaction since at this order the vacuum channel interaction is different.
It is remarkable however that they
 exactly coincide with the first two
terms found for corresponding correction in the vacuum channel. As a whole the
found expression is much simpler than the one for the vacuum channel,
which includes many more terms with a complicated dependence on $\lambda$.

\section{The  non-forward pair kernel in the octet colour state}
The bootstrap gives a relation between the gluon trajectory and its
pair interaction in the octet colour state. As mentioned this 
interaction is important for the  the odderon, where it enters in the 
combination (6).  In this section we shall obtain the 2nd order
correction to the kernel (6) using the found form of the function $\eta$
and Eqs (3) and (4).

It is convenient to present the kernel (6) in the form which explicitly shows
its infrared stability. We use an identity
\beq
\frac{1}{\ea\eb}=\frac{1}{\ea(\ea+\eb)}+\frac{1}{\eb(\ea+\eb)}
\eeq
to present the trajectory (3) in an equivalent form
\beq
\oq=-2\int
\frac{d^{2}q_{1}\eq}{(2\pi)^{2}\ea(\ea+\eb)},\ \ 
q=q_{1}+q_{2}.
\eeq
Then denoting $q_{11'}=q_1-q'_1$ for brevity and
combining the terms with $\ee$ in the denominator in the trajectories
and in the interaction, we obtain for the action of the  kernel (6) on an
arbitrary function $\psi(q_1)$
\[
K\psi=2\int\frac{d^2q'_1\ea}{(2\pi)^2\ee}\left(\frac{\psi(q'_1)}{\ec}-
\frac{\psi(q_1)}{\ec+\ee}\right)\]\beq+(1\leftrightarrow 2)
-2\int\frac{d^2q'_1\eq\psi(q'_1)}{(2\pi)^2\ec\ed}.
\eeq
As in the case of the standard BFKL equation, this form shows explicitly
that all singularities at small momenta $q'_1, q'_2$ and $q_{11'}$ cancel
and the kernel is infrared stable if function $\eq$ goes to zero
not essentially faster than $q^2$.

Now, preserving the infrared stability, we present in (44) function $\eta$ in  
the form (12) keeping terms up to the 2nd order in $\alpha_s$. For the 
2nd order contribution we obtain
\[
K^{(2)}\psi=6\alpha_s\int\frac{d^2q'_1q_1^2}{(2\pi)^2q_{11'}^2}
(\frac{\psi(q'_1)}{{q'_1}^2}(\xa-\xc-\xe)-\]\[
\frac{\psi(q_1)}{{q'_1}^2+q_{11'}^2}(\xa-\xe-
\frac{{q'_1}^2}{{q'_1}^2+q_{11'}^2}\xc-
\frac{q_{11'}^2}{{q'_1}^2+q_{11'}^2}\xe))\]
\beq+(1\leftrightarrow 2)
-6\alpha_s\int\frac{d^2q'_1q^2\psi(q'_1)}{(2\pi)^2{q'_1}^2{q'_2}^2}
(\xq-\xc-\xd).
\eeq
Presenting
\beq
\frac{{q'_1}^2}{{q'_1}^2+q_{11'}^2}=1-\frac{q_{11'}^2}{{q'_1}^2+q_{11'}^2},
\eeq
we separate from the first term a contribution
\beq
6\alpha_s\psi (q_1)\int\frac{d^2q'_1 q_1^2}{(2\pi)^2({q'_1}^2+q_{11'}^2)^2}
(\xc-\xe),
\eeq
 which is equal to zero, since it changes sign under the substitution
$q'_1\rightarrow q_1-q'_1$.

This brings us to a simple expression
\[
K^{(2)}\psi=6\alpha_s\int\frac{d^2q'_1q_1^2}{(2\pi)^2q_{11'}^2}
\left(\frac{\psi(q'_1)}{{q'_1}^2}-\frac{\psi(q_1)}{{q'_1}^2+q_{11'}^2}
\right)
(\xa-\xc-\xe)\]\beq+(1\leftrightarrow 2)
-6\alpha_s\int\frac{d^2q'_1q^2\psi(q'_1)}{(2\pi)^2{q'_1}^2{q'_2}^2}
(\xq-\xc-\xd).
\eeq

Now we have only to put our expression (14) for $\xq$ with the coefficients
given by (25) and (27). Since the kernel $K^{(2)}$ is infrared finite, all
terms in (14) proportional to $\epsilon$ or $\epsilon^2$ give no
contribution, and, as in the forward case, the result only depends on
$c_0=-\bar{g}^2B$ and $d=\bar{g}^2A$. It is also clear that the constant
term $c_0$ in $\xq$ just rescales the first order kernel $K^{(1)}$. So
in the end we finally obtain an explicit form for the action of $K^{(2)}$ as
\[
K^{(2)}\psi=\frac{3\alpha_s}{4\pi}\Bigl[ BK^{(1)}\psi+
6A\alpha_s\int\frac{d^2q'_1q_1^2}{(2\pi)^2q_{11'}^2}
\left(\frac{\psi(q'_1)}{{q'_1}^2}-\frac{\psi(q_1)}{{q'_1}^2+q_{11'}^2}
\right)
\ln\frac{q_1^2\mu^2}{{q'_1}^2q_{11'}^2}\]\beq+(1\leftrightarrow 2)
-6A\alpha_s\int\frac{d^2q'_1q^2\psi(q'_1)}{(2\pi)^2{q'_1}^2{q'_2}^2}
\ln\frac{q^2\mu^2}{{q'_1}^2{q'_2}^2} \Bigr].
\eeq

As in the forward case, one expects that acting on the
 eigenfunctions of the lowest order kernel, this 2nd order kernel
will give a shift for the eigenvalue plus some terms which come from the
running of the coupling. As mentioned, the latter should be the same as for the
non-forward kernel in the vacuum colour state. Unfortunately, although the found 
kernel does not look too complicated, we have not able to
find its action on the first order eigenfunctions up to now. 
Work in this direction is in progress.
\section{Conclusions}
We have found the 2nd order corrections to the interaction of two
reggeized gluons in the octet channel, using the bootstrap relation and
a specific ansatz to solve it proposed in [5]. This interaction is
important for the odderon. The relevant kernel for a pair
of gluons is found to be infrared finite and have a rather simple form.
For the forward case, apart from an expected term which describes the
running of the coupling, it contains a correction which coincides with the
first two terms found for the vacuum channel in [1,2].

The infrared stability of the found octet pair kernel,
if correct, implies that the three-body interaction in the odderon
has to be infrared finite by itself. In our approach
this stability, in fact, automatically
follows from the applied ansatz to satisfy the bootstrap relation.
We have no proof that this ansatz is unique. Calculation of the
three-body interaction for the odderon and study of its infrared
properties is thus essential to support our approach.  

After this work had been completed and published as a preprint [7] a
paper by V.Fadin {\it et al.} [8] appeared in which the quark
contribution to the interaction in the octet colour channel was directly
calculated
from the asymptotics of the relevant Feynman diagrams. It was
claimed in [8] that their results disgagreed with ours and that therefore
our bootstrap condition and the ansatz to solve it were wrong.
In fact this claim is absolutely unfounded: the quark contribution to
the interaction found by our method identically coincides with the
results of [8] (see Appendix), although the amount of labour necessary
to obtain it is incomparably smaller.
\section {Acknowledgments.}
The authors express their deep gratitude to Prof. G.Venturi for his
constant interest in this work. M.A.B. thanks the INFN  and Physics
Department of Bologna University for their hospitality and
financial support. G.P.V. gratefully acknowledges the financial support
from Italian M.U.R.S.T.

\section{Appendix}
The authors of  [8], in all probability, have not properly taken into
account that the unsymmetric kernel $V$ calculated in this paper
is a different  quantity as compared to their symmetric kernel $K_r$.
It is however straightforward to relate them. First one goes over from
$V$ to a symmetric kernel $W$ in the trivial metric (that is with the
integration measure $d^2q_1$) by expressing 
\beq
V(q,q_1,q'_1)=\sqrt{\frac{\ea\eb}{\ec\ed}}W(q.q_1,q'_1)
\eeq
On the other hand, the kernel $K_r$ of [8] is defined in the metric with 
an extra factor $q_1^2q_2^2$ in the denominator. Taking this into account
we obtain the desired relation
\beq
K_r(q,q_1,q'_1)=\sqrt{q_1^2q_2^2{q'_1}^2{q'_2}^2}
\sqrt{\frac{\ec\ed}{\ea\eb}}V(q,q_1,q'_1)
\eeq

Now we are going to show that, first, the quark contribution to the gluon
Regge trajectory has the form implied by our anzatz (3). We
shall determine the part of $\eq$ coming from this contribution.
Then using (4) and (51) we shall find the corresponding part of the
irreducible
kernel $K_r$ and compare it with the results of [3]. To simplify the
comparison with [8]
 we shall use the relevant expressions in the unrenormalized
form and for an arbitrary number of colours $N$. 

The part of the gluon trajectory which comes from quark is given by
the expression [8]
\beq
\omega_Q^{(2)}=2bq^2\int\frac{d^{D-2}q_1}{q_1^2q_2^2}(q^{2\epsilon}-
q_1^{2\epsilon}-q_2^{2\epsilon}),
\eeq
where
\beq
b=\frac{g^4NN_F{\rm \Gamma}(1-\epsilon){\rm \Gamma}^2(2+\epsilon)}
{(2\pi)^{D-1}(4\pi)^{2+\epsilon}\epsilon{\rm \Gamma}(4+2\epsilon)}.
\eeq
On the other hand using (3) and (12)
\beq
\omega^{(2)}(q)=-aq^2\int\frac{d^{D-2}q_1}{q_1^2q_2^2}(\xq-
\xa-\xb)
\eeq
where
\beq
a=\frac{g^2N}{2(2\pi)^{3+2\epsilon}}
\eeq
As we observe, the form of (52) follows this pattern. Comparing (52) and
(54)
we identify
\beq
\xi_Q(q)=-\frac{2b}{a}q^{2\epsilon}
\eeq

Now we pass to the irreducible kernel $K_r$. From (4) and (51) we
express it
via $\xq$ as
\[
K^{(2)}_r(q_1,q'_1)=\frac{1}{2}a\sqrt{q_1^2q_2^2{q'_1}^2{q'_2}^2}
\Big[\frac{1}{k^2}\sqrt{\frac{q_1^2{q'_2}^2}{{q'_1}^2q_2^2}}
(\xa+\xd-\xc-\xb-2\xe)
+\]\beq
\frac{1}{k^2}\sqrt{\frac{q_2^2{q'_1}^2}{{q'_2}^2q_1^2}}
(\xb+\xc-\xa-\xd-2\xe)
-\frac{q^2}{\sqrt{q_1^2q_2^2{q'_1}^2{q'_2}^2}}
(2\xq-\xa-\xb-\xc-\xd)\Big]
\eeq
Putting here the found $\xq$, Eq. (56), we obtain the quark contribution
\[
K^{Q}_r(q_1,q'_1)=-b
\Big[\frac{q_1^2{q'_2}^2}{k^2}
(q_1^{2\epsilon}+{q'_2}^{2\epsilon}-q_2^{2\epsilon}-{q'_1}^{2\epsilon}
-2k^{2\epsilon})+\]\beq
\frac{q_2^2{q'_1}^2}{k^2}
(q_2^{2\epsilon}+{q'_1}^{2\epsilon}-q_1^{2\epsilon}-{q'_2}^{2\epsilon}
-2k^{2\epsilon})
-q^2(2q^{2\epsilon}-q_1^{2\epsilon}-q_2^{2\epsilon}-{q'_1}^{2\epsilon}
-{q'_2}^{2\epsilon})
\Big]
\eeq
Comparing this expression with the one found in [8] (Eq. (47) of that
paper) we observe that they are identical. This means that our bootstrap
condition and the anzatz to solve it are valid at least for the quark
contribution in the next-to-leading order.

\section{References.}
\noi 1. V.S.Fadin, L.N.Lipatov, Phys. Lett. {\bf B429} (1998)127\\
2. G.Camici, M.Ciafaloni, Phys. Lett. {\bf B395} (1997) 118.\\
3. L.N.Lipatov, Yad. Fiz. {\bf 23} (1976) 642.\\
 J.Bartels, Nucl. Phys. {\bf B151} (1979) 293.\\
4. V.S.Fadin, R.Fiore, preprint BUDKERINP/98-54, UNICAL-TP 98/3, 
hep-ph/9807472\\
5. M.A.Braun, Phys. Lett. {\bf B345} (1995) 155; {\bf B348} (1995) 190.\\
6. V.S.Fadin, R.Fiore, M.I.Kotsky, Phys. Lett. {\bf B359} (1995) 181; 
{\bf B387} (1996) 593.\\
7. M.A.Braun and G.P.Vacca, hep-ph/9810454.\\
8. V.S.fadin, R.Fiore and A.Papa, hep-ph/9812456.\\

 \end{document}